\documentclass[conference]{IEEEtran}
\IEEEoverridecommandlockouts
\usepackage{cite}
\usepackage{amsmath,amssymb,amsfonts}
\usepackage{algorithmic}
\usepackage{graphicx}
\usepackage{textcomp}
\usepackage{xcolor}
\usepackage{booktabs}
\usepackage{times}
\usepackage{mathptmx}
\usepackage{helvet}
\usepackage{courier}
\usepackage{url}
\usepackage{subcaption}
\usepackage{hyphenat}
\usepackage{algorithm}
\usepackage{algorithmic}
\usepackage{hyperref}
\usepackage{fancyhdr}
\usepackage{msc}
\usepackage{makecell}
\usepackage{multirow}
\usepackage{multicol}
\def\BibTeX{{\rm B\kern-.05em{\sc i\kern-.025em b}\kern-.08em
    T\kern-.1667em\lower.7ex\hbox{E}\kern-.125emX}}
\begin{document}

\title{
Toward Fully Autonomous 6G Networks: AI-driven Operational Efficiency and Optimization
\\
\thanks{This work is supported by the Grant TRAINER-6G (PID2023-146748OB-I00) funded by MCIN/AEI/10.13039/501100011033 and by ERDF/EU.}
}

\author{
    \IEEEauthorblockN{
        David Reiss\IEEEauthorrefmark{1}, 
        Oriol Sallent\IEEEauthorrefmark{1},
        Miguel Catalan-Cid\IEEEauthorrefmark{2}, 
        Daniel Camps-Mur\IEEEauthorrefmark{2}, 
    }
    \IEEEauthorblockA{\IEEEauthorrefmark{1}UPC, Spain. \{david.reiss, jose.oriol.sallent\}@upc.edu}
    \IEEEauthorblockA{\IEEEauthorrefmark{2}i2CAT Foundation, Spain. \{david.reiss, miguel.catalan, daniel.camps\}@i2cat.net}
}

\maketitle

\begin{abstract}
Mobile networks evolution is characterized by a substantial increase in system complexity, driven by the need to accommodate a growing number of heterogeneous services on top of the digital infrastructure. This growth in service accommodation is expected to accelerate with the adoption of the Network as a Service (NaaS) paradigm, which has emerged as a promising approach to accelerate network innovation while enabling new revenue streams for operators. Although it is fundamental to abstract network capabilities for third-party developers, it poses significant challenges in terms of efficient network operation. To address this increased complexity, future mobile networks are envisioned to be inherently Artificial Intelligence (AI)-native. In particular, the integration of AI within the Radio Access Network (RAN) becomes a key enabler for optimizing operation, energy consumption, and autonomous network control. In this context, this research explores the convergence of AI-native RAN and  NaaS ecosystems to enable autonomous 6G RAN management. We propose an Agentic-based orchestration framework capable of interpreting intent-based policies. The proposed framework becomes key to integrate external NaaS requests with internal network management policies.
\end{abstract}

\section{Introduction}\label{sec:intro}

{The continuous evolution of mobile networks toward Beyond 5G (B5G) and Sixth Generation (6G) reveals an increase in system complexity. This complexity not only comes from a larger densification and virtualization of network infrastructure but also from the need to support a rapidly growing number of services and applications with diverse performance, reliability, and latency requirements. From a network management perspective, orchestrating such services necessitates the development and deployment of advanced mechanisms, such as network slicing or Quality on Demand (QoD), posing significant challenges in the operation of future networks. } 

{In this context, the adoption of Artificial Intelligence (AI) is expected to play a central role in addressing these operational challenges. 6G networks are anticipated to be inherently AI-native, endowing the digital infrastructure with autonomy and self-optimizable capabilities. Particularly, the integration of AI within the Radio Access Network (RAN) becomes crucial to foster innovation in areas such as intelligent resource management, energy efficiency, and autonomous network control. This integration is supported by the Open RAN (O-RAN) architecture \cite{o-ran-overview}, which introduces standardized open interfaces and deploys the so-called RAN Intelligent Controllers (RICs), facilitating the seamless integration of AI-driven control loops within the RAN. Building upon these architectural foundations, the AI-RAN Alliance \cite{ai-ran-alliance} brings together telco industry leaders to detail the role of AI in next-generation systems. The initiative identifies how AI can be leveraged to improve operational efficiency, to promote the execution of AI and generative AI applications within the RAN infrastructure, and to unlock new monetization opportunities, envisioning the network as shared infrastructure where RAN and non-RAN related AI workloads can be accommodated. }

{In this regard, the Network as a Service (NaaS) paradigm promotes the external consumption of network services by abstracting and exposing network capabilities, thereby enabling third-party applications and service providers to dynamically request, discover, and consume network resources on demand, emerging as a complementary approach to enabling new revenue streams for telco operators. This is reinforced by multiple initiatives, from which we highlight the Linux Foundation's CAMARA project, found within the scope of the GSMA Open Gateway collaborative framework \cite{open-gateway}. CAMARA is an open-source project that unites Mobile Network Operators (MNOs) to design an ecosystem where network capabilities can be exposed through northbound, developer-friendly, Application Programming Interfaces (APIs). By enabling third-party applications to seamlessly consume network services, CAMARA fosters the adoption of the NaaS paradigm. Within this context, the CAMARA's roadmap acknowledges the need for intelligence behind these APIs, and plans to evolving towards the so-called \textit{Agentic APIs}. These evolved interfaces enable not just the exploration of AI-driven use cases within the CAMARA ecosystem but also ensures that APIs transition towards AI driven interactions rather than manual programmatic calls. }

{AI Agents are expected to become a native component of 6G networks, enabling intent-driven orchestration and becoming crucial to deliver the adaptive capabilities necessary to manage these highly dynamic environments. For instance, authors in \cite{maestro-framework} present MAESTRO, a framework that utilizes LLM-based agents to automate negotiations among stakeholders. Unlike traditional optimization methods, this approach leverages the semantic reasoning of LLMs to interpret high-level intents from business operators (e.g., \textit{keep low OPEX and a minimum SLA of 20 Mbps}) to ground them in technical reality through specialized optimizer units. This state-of-the-art design addresses critical challenges in 6G orchestration to ensure conflict management and fair distribution of shared network resources.}

{Building on the above discussion, this thesis focuses on achieving an efficient RAN management through AI-guided automation. During the initial stage, we have demonstrated how AI-driven mechanisms can effectively improve the operational efficiency of the RAN. Moreover, these mechanisms have been integrated into an O-RAN–compliant emulation framework that enables the deployment and validation of AI-driven xApps/rApps as it would happen in an O-RAN live deployment. In line with recent trends, we envision that the adoption of the NaaS paradigm will become a crucial aspect of future network infrastructures, where external network requests will coexist with internal management policies. Accordingly, this paper further explores research directions devoted to incorporate NaaS workflows within the 6G RAN management ecosystem. After identifying the main challenges associated with the NaaS adoption and detailing the key steps for a reliable execution, the final part of the paper details an orchestration framework on which network AI Agents are capable of interpreting and exchanging policies expressed in Natural Language (NL), fostering an autonomous, intent-driven 6G RAN operation. }

{The paper is structured as follows. The next section reviews the work completed across the initial stage of this thesis, and highlights key management challenges in future networks. After that, we present the future thesis framework. Finally, in Section \ref{sec:conclusions}, we present the conclusions and the open questions.}

\section{Research topics and challenges}\label{sec:use-case}

{In this section we detail the current stage of the PhD and the future research directions. Subsection A describes how AI techniques can be leveraged to improve the efficiency of the RAN and introduces an O-RAN–compliant framework developed for the evaluation of AI-driven optimization xApps/rApps. Subsection B then identifies the key challenges that appear in NaaS ecosystems, not only in terms of RAN management but also in feasibility assessments.}

\subsection{Current stage: AI for RAN optimization}

\begin{figure*}[t]
    \centering
    \includegraphics[width=0.98\textwidth]{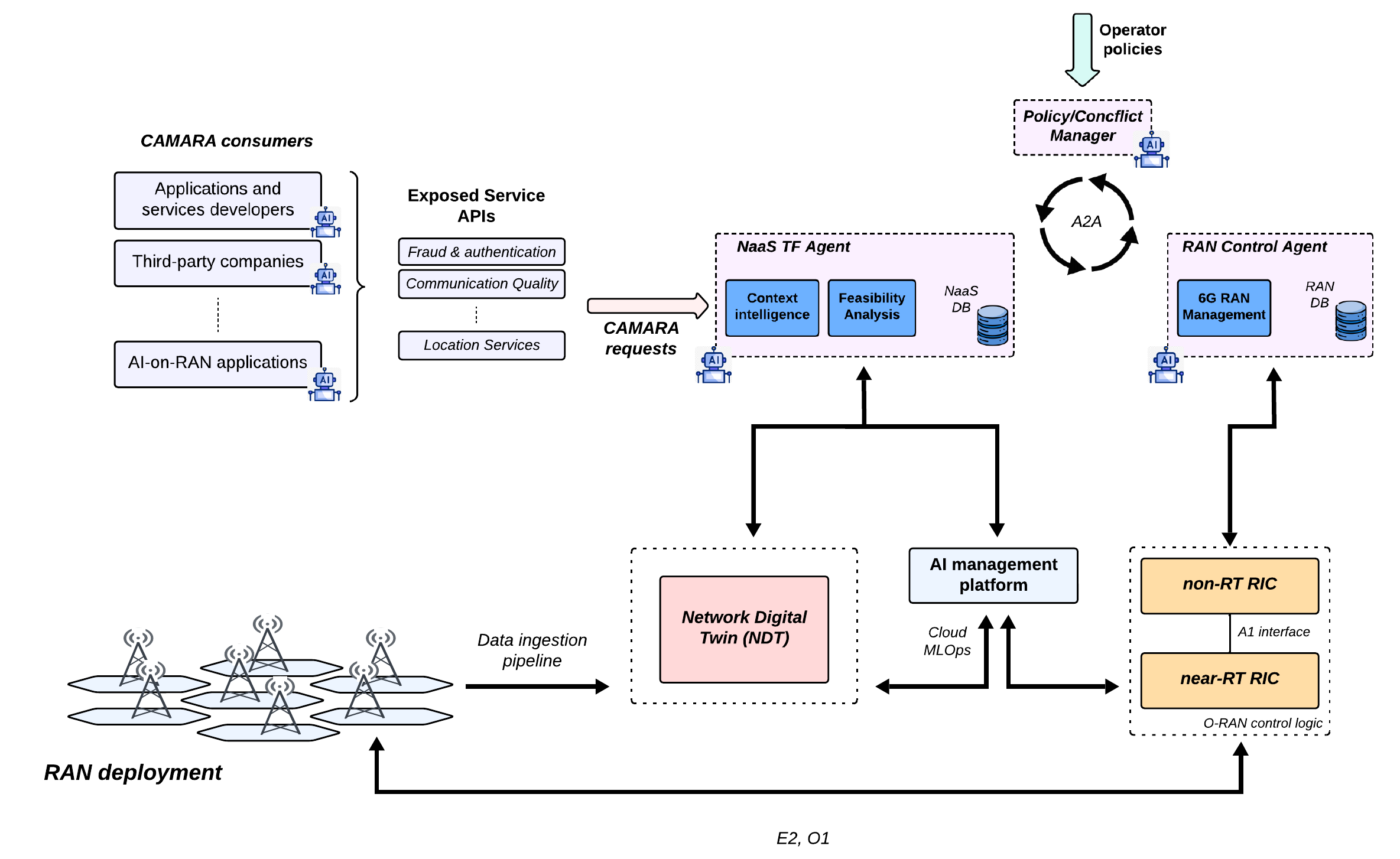}
    \caption{High-level description of future thesis framework.}
    \label{fig:high-level}
\end{figure*}

{The initial stage of this PhD has primary focused on improving the operational efficiency of current 5G networks. Analyzing a real cellular network dataset provided by an European MNO revealed that the 5G NSA deployment suffered from over-provisioning in specific areas, which is commonly observed during transitions to new mobile generations. To mitigate unnecessary energy consumption, we developed AI-driven mechanisms which were not only capable of identifying 5G low-utilization scenarios but also incorporated QoS-aware mechanisms, providing tools for MNOs to balance the achievable savings and the end users service requirements, \cite{mswim-paper}. Remarkably, those were trained using standardized Performance Measurement (PM) counters provided by the MNO's infrastructure provider, ensuring robustness and facilitating the future adoption of such strategies in 6G deployments. }

{Evaluating these AI-driven mechanisms is a crucial step to validate the correct functioning prior to its commercial deployment. To this end, we have developed an O-RAN emulation framework which is composed of three main blocks \cite{demobegreen}. First, a decoupled AI management platform is used to train, update, and expose AI/ML models through inference endpoints that can be later invoked by RIC-hosted xApps/rApps. Second, the O-RAN control module is composed of both the non-RT RIC and the near-RT RIC, fully aligned with the O-RAN Software Community (OSC) guidelines and implementing the full set of O-RAN standard interfaces, APIs, and terminations. Finally, the emulation block is the Keysight RIC Test, which enables the exchange of telemetry with RICs through E2 and O1 interfaces. In addition, a data ingestion pipeline has been developed to tailor the emulation configuration parameters to the dataset provided by the MNO. The full framework demonstrates several functionalities, such as the exposure of KPM datasets through the E2 interface, the enforcement of AI-driven management policies, or the high-accurate emulation of real RAN scenarios. Together, these functionalities are developed to transition toward an O-RAN Network Digital Twin (NDT), which is expected to play a key role in future thesis stages. As detailed in the following section, the reliable assessment of external NaaS workflows will be strictly linked to the NDT.    }

\subsection{Future directions: AI for RAN management in NaaS ecosystems}

{As highlighted in the introduction, the network is evolving to incorporate new paradigms, such as the NaaS. In this context, the CAMARA initiative focuses on exposing low-level network capabilities in a customer-facing environment where no telco expertise is required. API consumers are therefore able to explore, discover, and request prioritized network resources in a programmatic manner. These end consumers can be of any type, from third-party companies, to AI-driven applications and services. Although CAMARA is developing a huge set of API families, a strong interest from telco operators is identified with respect to the programmable connectivity interfaces \cite{Juniper-research}. These type of interfaces are addressed within the Communication and Quality API family, involving use cases such as QoS booking, QoD, etc. }

{Note that, these type of APIs will typically require MNOs to set specific network slices or dedicate a certain amount of resources to prioritize API consumers and therefore, implementing such requests is a complicated task. Although CAMARA defines the transformation functions (TFs), which are the custom logic below the exposed service APIs, its implementation is left undefined. In addition, we consider that responding to API requests should be based on a strong feasibility analysis executed prior to its implementation, as several network management conflicts may arise. In this context, the future objective of this thesis is to define methodologies to respond and manage the API requests, by transforming the high-level service API calls into detailed RAN-level, telco-expert assessments. Notably, this becomes key to ensure the autonomous management of NaaS workflows within the RAN infrastructure. Concretely, we detail two main tasks associated with the viability assessment of CAMARA requests:  }

\begin{itemize}
    \item \textbf{Network context intelligence:}{ leverages AI-driven mechanisms to assess time-critic API calls, such as a QoD request for a specific area over the next hour. In such scenarios, combining network information (e.g., current RAN status or historical cell utilization) and intent-based runtime descriptions (e.g., the occurrence of a public event, or weather information) are crucial to extract the network context of the affected areas. The developed AI management platform provides a suitable environment for the execution of these rapid, context-aware assessments. }
    \item \textbf{Long-term feasibility assessment:} {provides feasibility assessments for non-time-critic API requests, such as QoS bookings for a specific area scheduled days or weeks in advance. It evaluates RAN scenarios to identify potential risks, including predicted cell saturation and QoE degradation of other network users. Based on these assessments, mitigation actions can be planned in advance, such as adjusting transmit power of neighboring cells or activating additional coverage carriers. Due to its realistic scenario reproduction capabilities, the already-developed O-RAN emulation framework is a suitable environment to execute and evaluate this type of tasks. As previously mentioned, transitioning to O-RAN NDTs \cite{dts}, is key to provide robust and reliable assessments. }
\end{itemize}

{The final task of the thesis focuses on the API execution. The introduction of NaaS workflows adds a new layer of complexity to the RAN management, as prioritized and non-prioritized users are expected to coexist more frequently within the cellular network. In particular, scenarios in which prioritized CAMARA consumers do not fully utilize their reserved resources may lead to suboptimal spectrum usage and overall operational inefficiencies. Addressing this challenge requires the design of dynamic scheduling and resource allocation mechanisms that can adapt to the characteristics, duration, and criticality of each CAMARA service. Such mechanisms are essential to enable efficient resource sharing while preserving service guarantees, ultimately ensuring the operational efficiency and scalability of 6G RANs under NaaS adoption.}
\section{Proposed Agentic-based architectural framework}\label{sec:solution-design}

The overall thesis framework is outlined in Fig. \ref{fig:high-level}. Note that, this represents an initial design on which many of its elements can be modified or re-designed. In line with current trends, the previously described tasks are mapped onto three main AI agents. These AI agents will be crucial to merge NaaS workflows with an efficient intent-driven RAN management:

\begin{itemize}
    \item \textbf{RAN Control Agent: }{this agent is responsible for the 6G RAN management. It is capable of executing Self-Organizing Network (SON) policies, such as, decreasing energy consumption or optimizing coverage for specific areas. In addition, this agent can also communicate with the NaaS TF Agent to coordinate accepted CAMARA requests. Similarly as in \cite{agentran}, we envision this agent capable of interacting with lower-layer agents, dedicated to radio power control, scheduling management, etc. In addition, it uses the RAN database (DB) to collect future RAN actions (associated to CAMARA or SON). }
    \item \textbf{NaaS TF Agent: }{this agent manages external NaaS consumption workflows. Its primary role is to conduct feasibility assessment of requested CAMARA calls, as well as establishing its execution requirements, therefore implementing the TFs. This agent leverages the context intelligence, and feasibility assessment tools, and moreover is able to communicate with the RAN Control Agent to get network status as well as to communicate accepted requests execution. Moreover, it is capable of communicating with the Policy/Conflict Manager to check conflicts with operator policies. In addition, it collects pending and accepted requests on its NaaS DB, therefore able to check conflicts on overlapping requests. }
    \item \textbf{Policy/Conflict Manager: }{this agent is responsible for interpreting operator-defined policies either to enforce SON control actions or externally triggered CAMARA requests. Its primary role is to manage conflicts that may arise between internal RAN optimization objectives and external CAMARA API demands, particularly when such actions overlap in time, geographic area, or resource utilization. By leveraging intent-based policies, this agent ensures coherent and policy-compliant RAN management. }
\end{itemize}

{Communication among these three agents would be governed by the Agent-to-Agent (A2A) protocol, currently under development by Google. This protocol standardizes inter-agent interactions by enabling the exchange of information through exposed APIs using NL, or structured JSON formats. A simple interaction example is illustrated in Algorithm \ref{alg:qos_booking}.}

\begin{algorithm}[htbp]
\caption{Example of Agents interaction}
\label{alg:qos_booking}
\begin{algorithmic}[1]
\STATE \textbf{Operator} $\rightarrow$ \textbf{Policy/Conflict Manager (PCM)}: prioritize revenue streams over SON policies.
\STATE \textbf{CAMARA} $\rightarrow$ \textbf{NaaS TF Agent}: QoS booking: $15$ devices, GBR $10$ Mbps, duration $1$h, location $X$, short time-to-response.
\STATE \textbf{NaaS TF Agent}: checks affected cell $\rightarrow$ starts context intelligence (short time-to-response).
\IF{non-feasible booking}
    \STATE \textbf{NaaS TF Agent} $\rightarrow$ \textbf{CAMARA}: reject
    \RETURN
    \ELSE
    \STATE \textbf{NaaS TF Agent} $\rightarrow$ \textbf{PCM}: feasible request, check operator policies.
    \STATE \textbf{PCM} $\rightarrow$ \textbf{NaaS TF Agent}: according to operator policy revenue has priority, request accepted.
    \STATE \textbf{NaaS TF Agent} $\rightarrow$ \textbf{RAN Control (RC) Agent}: create a slice to allocate 15 UEs with a GBR of 10 Mbps in cell $Y$ for 1h. It has priority over SON policies.
    \IF{cell $Y$ is in energy-saving mode}
        \STATE \textbf{RC Agent}: re-activate cell $Y$.
    \ENDIF
    \STATE \textbf{RC Agent} configures a dedicated slice.
    \STATE \textbf{RC Agent} $\rightarrow$ \textbf{NaaS TF Agent} $\rightarrow$ \textbf{CAMARA}: QoS booking request accepted.
\ENDIF
\end{algorithmic}
\end{algorithm}

\section{Conclusions}\label{sec:conclusions}

{This paper has explored the key challenges to integrate external NaaS workflows within the 6G RAN management. The completed work corresponding to the initial stage of the thesis has been discussed, demonstrating how operational efficiency can be improved through AI-powered algorithms and validated in an O-RAN compliant framework, which already serves as an initial NDT for future thesis stages. Furthermore, in alignment with recent industry trends, we have detailed an environment where AI Agents embed the necessary tasks devoted to ensure a reliable NaaS adoption within the 6G RAN management. }

{In this respect, we identify two open challenges. The first challenge concerns conflict management among AI agents. While the illustrative example considers a simplified conflict scenario in which the operator policy allows a straightforward resolution, more complex situations (such as dynamically evolving policies, or overlapping objectives) remain largely unexplored. Addressing these scenarios requires the development of advanced coordination, and negotiation mechanisms among agents to ensure consistent decision making. The second challenge relates to the design of the NDT. The NDT must balance simplicity and robustness, enabling lightweight and efficient interaction with AI agents while still providing robust and reliable assessments of network behavior. Relying on a full O-RAN stack for such evaluations may introduce prohibitive overheads in terms of scalability and computational cost, motivating the exploration of adaptive NDT designs tailored to the requirements of AI-driven RAN management. }

\bibliographystyle{IEEEtran}
\bibliography{references.bib}

\end{document}